\documentclass[12pt,leqno]{article}
\usepackage{amsmath,amsthm,amssymb}

\def\init{\setcounter{equation}{0}}
\newtheorem{theorem}{Theorem}[section]

\newcommand{\Z}{\mathbb{Z}}

\newcommand{\e}{{\varepsilon}}

\newcommand{\rw}{\rightarrow}

\usepackage{tikz}
\usetikzlibrary{decorations.pathreplacing}

\usepackage{verbatim}

\title{Hawking type radiation from  acoustic    black holes  with   time-dependent  metric}

\author{G.Eskin, \ \ \  Department of Mathematics, UCLA,\\ Los Angeles,
CA 90095-1555, USA. \ E-mail: eskin@math.ucla.edu
}

\begin{document}

\maketitle

\begin{abstract}
  We consider  a time-dependent acoustic   metric  in 2+1  dimensions  having  a black hole 
  and we study  the Hawking type  radiation   from such black hole.
  
  We construct  eikonals of special form  with the support  close to the  black  hole.  We  compute the average number
  of created  particles  where  the average  is taken  with  respect  to the  Unruh  type  vacuum.  
 
\end{abstract}

\section{Introduction}
\init
In this paper  we study the rotating  acoustic  black  holes  when  the metric  is 
\begin{equation}															\label{eq:1.1}
\Big(dx-\frac{A(x_0)}{|x|}\hat x dx_0\Big)^2,
\end{equation}
where  $x=(x_1,x_2), \hat x= \frac{x}{|x|}, A(x_0)<0, A(x_0)\rw A(+\infty)$  when  $x_0\rw +\infty,\ A(x_0)\rw A(-\infty)$  when  $x_0\rw -\infty$.

The wave  operator  $\Box_g$,  corresponding to the metric (\ref{eq:1.1}),   has the form 
\begin{equation}															\label{eq:1.2}
\Box_g u=  \Big(\frac{\partial}{\partial x_0} +\frac{A(x_0)}{\rho}\frac{\partial}{\partial\rho}\Big)^2 - \frac{\partial^2}{\partial \rho^2}
-\frac{1}{\rho^2}\frac{\partial^2}{\partial\varphi^2},
\end{equation}
where  $(\rho,\varphi)$  are polar coordinates.

It  was  shown  in [1]  that  metric  (\ref{eq:1.1})    has  a black hole  
$B=\{0<|x|<\rho^*(x_0)\}$  when  $A<0$.   In this paper  we  study  the Hawking  type  radiation  from such  black  holes.

In  \S  2  we introduce elements  of the quantum fields  theory on curved spacetimes following mostly  the lecture notes  of 
T.Jacobson  [6]  (see  also [4]).
We define  the number  operator  of  created particles  and we use the Unruh type  vacuum  state (cf.  [7], [2])  to  compute  the 
average number  of created particles.

In \S 3  we  review   (cf.  [1])  the black holes  for the rotating time-dependent acoustic  metrics.   In \S 4  we embark  on the computation  of
the average  number of particles  created by some special  wave   packets  that are   similar to the wave packets  used in the case  
of stationary rotating  acoustic  metrics (cf.  [2]).

More  precisely,   we take
$$
C_0(\rho,x_0,\alpha)=\frac{1}{\sqrt \rho} e^{i\alpha \ln(\sigma(\rho,x_0)-\sigma_*)-a(\sigma(\rho,x_0)-\sigma_*)}
(\sigma(\rho,x_0)-\sigma_*)^\e  \theta(\sigma(\rho,x_0) -\sigma_*),
$$
where  $\alpha>0$  is  a constant,     
$\sigma(\rho,x_0)$   is  the solution  of  
$$
\frac{\partial\sigma}{\partial x_0}+\Big(\frac{A(x_0)}{\rho}+1\Big)\frac{\partial\sigma}{\partial\rho}=0,\ \ \sigma\Big|_{x_0=0}=\rho,
$$
and $\sigma_*=\rho^*(x_0)$  when  $x_0=0$.

Note that  the cases  of black  holes  in  the  present  paper  (as in  the paper  [2]  that  treats  the  stationary  acoustic metric
with variable velocity)  are not accessible   by the methods of previous works  of S.Hawking,  T.Jacobson  and many  others.   We   use  
a new approach  in a new situation.  Thus we  introduce  a splitting   on positive and negative modes,  
and we modify the definition of  Unruh vacuum  state.   We construct  appropriate wave packets  to enhance the Hawking radiation.

The  guiding  motivation  for our new  definitions  and constructions is  that  they are natural,
  and in the case  of overlapping   with the  previous results  they give
  the same  (modulo greybody  factors)  exponential   expression   for the  Hawking  radiation.

\section{Elements  of quantum  field  theory on a curved  spacetime}
\init

We  introduce  elements  of quantum  fields theory on curved  spacetime   following  the  lecture  notes  of  T.Jacobson  [6].

Let  $<u,v>$  be the  Klein-Gordon (KG)  inner  product  of  $u$  and  $v$,   i.e.
\begin{equation}																\label{eq:2.1}
<u,v>=i\int\limits_0^\infty\int\limits_0^{2\pi}\Big[\Big(\overline u\frac{\partial v}{\partial x_0}-\frac{\partial\overline u}{\partial  x_0}v\Big)
+\frac{A(x_0)}{\rho}\Big(\overline u\frac{\partial v}{\partial\rho}-\frac{\partial\overline u}{\partial \rho}v\Big)\Big]
\rho d\rho d\varphi.
\end{equation}
Here  $x_0$  is fixed.  When  $u,v$  are  the solutions of the wave equation  (\ref{eq:1.2})  and  the  metric
is independent of $x_0$ then  $<u,v>$  is  independent   of  $x_0$
(cf.  [6]).

Denote   by  $f_k^+(x_0,\rho,\varphi),f_{-k}^-(x_0,\rho,\varphi)$   the solutions  of  (\ref{eq:1.2})  with the following  initial  conditions  (cf.  [2]):
\begin{align}																	\label{eq:2.2}
&f_k^\pm(0,\rho,\varphi)=\gamma_0 e^{i\rho\eta_\rho+im\varphi},
\\
\nonumber
&\frac{\partial f^\pm(0,\rho,\varphi)}{\partial  x_0}=i\lambda^\mp(k)\gamma_0 e^{i\rho\eta_\rho+im\varphi},
\end{align}
where  $k=(\eta_\rho,m),  m\in \Z, \gamma_0=\frac{1}{2\pi\sqrt 2}\,\frac{1}{\sqrt \rho(\eta_\rho^2+1)^{\frac{1}{4}}}$  is a normalizing factor,
\begin{equation}																\label{eq:2.3}
\lambda^\pm(k)=-\frac{A(x_0)}{\rho}\eta_\rho\pm \sqrt{\eta_\rho^2+1}.
\end{equation}
Note that
 \begin{equation}																\label{eq:2.4}
 \overline{f_k^+(x_0,\rho,\varphi)}=f_{-k}^-(x_0,\rho,\varphi).
 \end{equation} 
As in [2]   one  can  show  that
\begin{align}																	\label{eq:2.5}
&<f_k^+(0,\rho,\varphi),f_{k'}^+(0,\rho,\varphi)>=\delta(\eta_\rho-\eta_\rho')\delta_{mm'},
\\
\nonumber
&
<f_{-k}^-(0,\rho,\varphi),f_{-k'}^-(0,\rho,\varphi)>=-\delta(\eta_\rho-\eta_\rho')\delta_{mm'},\ \ 
<f_k^+,f_{k'}^->=0,\ \ \forall k,\forall k',
\end{align}
where
\begin{equation}																\label{eq:2.6}
k=(\eta_\rho,m),k'=(\eta_\rho',m').
\end{equation}
Thus the following theorem  holds:
\begin{theorem}																\label{theo:2.1}
Let $\{ f_k^+(0,\rho,\varphi),f_{-k}^-(0,\rho,\varphi)\}$  form  an  ``orthogonal"  basis  with  respect  to KG  inner  product.
Let  $C(x_0,\rho,\varphi)$  be smooth.  Then  $C(x_0,\rho,\varphi)$  can be expanded   in $\{f_k^+,f_{-k}^-\}$:
\begin{equation}																\label{eq:2.7}
C(x_0,\rho,\varphi)=\sum_{m=-\infty}^\infty\int\limits_{-\infty}^\infty C_k^+f_k^+(x_0,\rho,\varphi)d\eta_\rho
+\sum_{m=-\infty}^\infty\int\limits_{-\infty}^\infty C_{-k}^-f_{-k}^-(x_0,\rho,\varphi)d\eta_\rho,
\end{equation}
where
\begin{equation}																\label{eq:2.8}
C_k^+=<f_k^+,C>\big|_{x_0=0},\ \  C_{-k}^-=-<f_{-k}^-,C>\big|_{x_0=0}.
\end{equation}
Note that  the KG  inner product  is taken  for  $x_0=0$.
\end{theorem}
Let  $\Phi(x_0,\rho,\varphi)$  be  the wave  operator  (cf.  [6]):
\begin{equation}   																\label{eq:2.9}
\Box_g\Phi =0.
\end{equation}
Expanding  $\Phi$  in the basis  $\{ f_k^+,f_{-k}^-\}$  we get
\begin{equation}																\label{eq:2.10}
\Phi=\sum_{m=-\infty}^\infty\int\limits_{-\infty}^\infty\big(\alpha_k^+f_k^+(x_0,\rho,\varphi)+\alpha_{-k}^-f_{-k}^-(x_0,\rho,\varphi)\big)d\eta_\rho,
\end{equation}
where  $\alpha_k^+,\alpha_{-k}^-$   are  called   the   annihilation  and creation  operators
\begin{equation}																\label{eq:2.11}
(\alpha_k^+)^*=\alpha_{-k}^-,\ \ \alpha_k^+=<f_k^+,\Phi>\big|_{x_0=0},\ \ \alpha_{-k}^-=-<f_{-k}^-,\Phi>\big|_{x_0=0}.
\end{equation}
These operators  satisfy  the following  commutation  relation  (cf.  [6]):
\begin{equation}																\label{eq:2.12}
[\alpha_k^+,\alpha_{-k'}^-]=\delta(\eta_\rho-\eta_\rho')\delta_{mm'}I,\ \ \ [\alpha_k^+,\alpha_{k'}^+]=0,
\end{equation}
\begin{equation}																\label{eq:2.13}
[\alpha_k^-,\alpha_{k'}^-]=0,\ \ k=(\eta_\rho,m),\ k'=(\eta_\rho,m'),
\end{equation}
$I$  is   the identity  operator.

To introduce   the Unruh  type  vacuum space  we need to split  $f_k^+(x_0,\rho,\varphi)$  and  $f_{-k}^-(x_0,\rho,\varphi)$  into two parts:
\begin{equation}																\label{eq:2.14}
f_k^{++}=f_k^+ \theta(\eta_\rho),\ \ f_k^{+-}=f_k^+(1-\theta(\eta_\rho)),
\end{equation}
where
$\theta(\eta_\rho)=1$  when  $\eta_\rho>0,\ \theta=0$  when  $\eta_\rho<0$.
Analogously
\begin{equation}																\label{eq:2.15}
f_{-k}^{-+}=f_{-k}^- \theta(\eta_\rho),\ \ f_{-k}^{--}=f_{-k}^-(1-\theta(\eta_\rho)),
\end{equation}
\begin{align}																       \label{eq:2.16}
&\alpha_k^{++}=\alpha_k^+\theta(\eta_\rho),\ \alpha_k^{+-}=\alpha_k^+(1-\theta(\eta_\rho)),
\\ \nonumber
& \alpha_{-k}^{-+}=\alpha_{-k}^-\theta(\eta_\rho),\ 
\alpha_{-k}^{--}=\alpha_{-k}^-(1-\theta(\eta_\eta)).
\end{align}
Using  (\ref{eq:2.15}),   (\ref{eq:2.16}),   we have
\begin{equation}																\label{eq:2.17}
\Phi=\sum_{m=-\infty}^\infty\int\limits_{-\infty}^\infty\big(\alpha_k^{++}f_k^{++}+\alpha_k^{+-}f_k^{+-}
+\alpha_{-k}^{-+}f_{-k}^{-+} +\alpha_{-k}^{--}f_{-k}^{--}\big)d\eta_\rho.
\end{equation}
Operator  $N$  of the number  of the particles  created  by    packet   $C(x_0,\rho,\varphi)$  is  defined  as  (cf. [6], [4]):
\begin{equation}																\label{eq:2.18}
N= <C,\Phi>^* <C,\Phi>.
\end{equation}         
We shall  define  the  Unruh  type  vacuum  state  $|\Psi\rangle$   by the conditions
\begin{equation}																\label{eq:2.19}
\alpha_k^{++}|\Psi\rangle =0,\ \ \alpha_{-k}^{--}|\Psi\rangle=0,  \ \ \forall k.
\end{equation}
The average  number  of created  particles  is
\begin{equation}																\label{eq:2.20}
\langle\Psi | N|\Psi\rangle.
\end{equation}
Analogously  to [2]  we can prove
\begin{theorem}																\label{theo:2.2}
The  average  number  of particles  created  by the  wave packet $C$  is given  by
\begin{equation}																\label{eq:2.21}
\langle\Psi | N|\Psi\rangle =\sum_{m=-\infty}^\infty\int\limits_{-\infty}^\infty\big(-|\overline{C^{+-}(k)}|^2+|C^{-+}(k)|^2\big)d\eta_\rho.
\end{equation}
\end{theorem}
{\bf Proof:}
Since   $\alpha_k^{++}|\Psi\rangle =0,\ \alpha_{-k}^{--}|\Psi\rangle=0$  we have  
\begin{equation}																\label{eq:2.22}
<C,\Phi>|\Psi\rangle=\sum_{m=-\infty}^\infty\int\limits_{-\infty}^\infty\big(\overline{C^{+-}(k)}\alpha_k^{+-}|\Psi\rangle-
\overline{C^{-+}(k)}\alpha_{-k}^{-+}|\Psi\rangle \big)d\eta_\rho,
\end{equation}
where $C^{+-}(k)=<f_k^{+-},C>,\ C^{-+}(k)= -<f_{-k}^{-+},C>$.  Since
\begin{equation}																\label{eq:2.23}
\langle\Psi|(\alpha_k^{++})^*=0,\ \ \langle\Psi|(\alpha_{-k}^{--})^*=0,
\end{equation}
we have
\begin{equation}																\label{eq:2.24}
\langle\Psi |(C,\Phi)^*=
\sum_{m=-\infty}^\infty\int\limits_{-\infty}^\infty \langle\Psi | C^{+-}(\alpha_k^{+-})^*-
\langle\Psi | C^{-+}(\alpha_k^{-+})^*d\eta_\rho.
\end{equation}
Therefore,  as  in  [2],  we get
 \begin{equation}																\label{eq:2.25}
\langle\Psi | N|\Psi\rangle =\sum_{m=-\infty}^\infty\int\limits_{-\infty}^\infty\big(-|C^{+-}(k)|^2+|C^{-+}(k)|^2\big)d\eta_\rho.
\end{equation} 

\section{Black  holes  in rotating  acoustic  time-dependent metric}
\init
Consider  a rotating  fluid  with the velocity  
\begin{equation}																\label{eq:3.1}
\vec v=\frac{A(x_0)}{\rho}\hat x,
\end{equation}
where
$\hat x=\frac{(x_1,x_2)}{|x|},\rho=|x|,x_0$   is   the time  coordinate,  $A(x_0)\in  C^\infty,  A(x_0)<0$  for all  
$x_0,\ \lim_{x_0\rw +\infty} A(x_0)=A(+\infty),  \ \lim_{x_0\rw -\infty} A(x_0)=A(-\infty).$
We assume   that  there  is no dependence  on $\varphi$,  where  $(\rho,\varphi)$  are polar  coordinates.   The  acoustic  metric  
associated  with this  flow  has  the  form  (cf. [9])
\begin{equation}																\label{eq:3.2}
\Big( dx -\frac{A(x_0)}{|x|}\hat x dx_0\Big)^2,\ \ x=(x_1,x_2),
\end{equation}
and  the corresponding  wave  equation  is
\begin{equation}																\label{eq:3.3}
\Box_g u \equiv \Big(\frac{\partial}{\partial x_0}+\frac{A(x_0)}{\rho}\frac{\partial}{\partial\rho}\Big)^2 u -\frac{\partial^2 u}{\partial \rho^2}
-\frac{1}{\rho^2}\frac{\partial^2 u}{\partial\varphi^2}=0.
\end{equation}
The Hamiltonian  (symbol)  of  $\Box_g$  is 
$$
H=\Big(\xi_0+\frac{A(x_0)}{\rho}\xi_\rho\Big)^2-\xi_\rho^2-\frac{1}{\rho^2}\xi_\varphi^2.
$$
The case  of time-dependent acoustic  metric  was  studied in  [1].  It was  proven  there that  when  $A(x_0)<0$  for all $x_0$  
 there exists  a black hole  $B=\{0<\rho<\rho^*(x_0)\}$  where  $\rho^*(x_0)$   is  the solution  of
\begin{equation}																\label{eq:3.4}
\frac{d\rho}{d x_0}=  \frac{A(x_0)}{\rho}+1,
\end{equation}
with some initial data
\begin{equation}	 															\label{eq:3.5}
\rho^*(0)=\sigma^*.
\end{equation}
Note  (cf.  [1])  that  all solutions  $\rho=\rho(x_0)$   of  (\ref{eq:3.4})
tends  to  $A(-\infty)$  when  $x_0\rw -\infty$.   When  $\rho(o)=\sigma >\sigma^*$  then  the solution  $\rho=\rho(x_0)$  tends  to 
$+\infty$  when  $x_0\rw  +\infty$  and  when  $\rho(0)=\sigma'<\sigma^*$  then  the solution  $\rho=\rho(x_0)$  tends  to zero  when  
$x_0\rw +\infty$.   Thus the solution  $\rho=\rho*(x_0)$  separates  the solutions of  (\ref{eq:3.4})  that  tend  to $+\infty$  when  $x_0\rw +\infty$
from the solutions of  (\ref{eq:3.4})  that tend  to $0$  when  $x_0\rw +\infty$.

Note  that  $\rho^*(x_0)\rw |A(-\infty)|$  when  $x_0\rw -\infty$  and  $\rho^*(x_0)\rw  |A(+\infty)|$  when  $x_0\rw  +\infty$.

In the  next  section  we shall study  the Hawking  type  radiation  from  the  acoustic  black hole  $B=\{0<\rho <\rho^*(x_0)\}$.

\section{Hawking type  radiation  from  rotating  acoustic  black hole}
\init

We shall  use  wave  packet  $C(x_0,\rho)$  independent  of  $\varphi$.

Let  $f_k^\pm(x_0,\rho,\varphi)$  be the same  as  in  (\ref{eq:2.2}).
Since  $C(x_0,\rho)$  is independent  of $\varphi$  we  have
\begin{equation}																\label{eq:4.1}
\int\limits_0^\infty\int\limits_0^{2\pi}
f_{\eta_\rho,m}^\pm(x_0,\rho,\varphi)C(x_0,\rho)\rho d\rho d\varphi=
\begin{cases}
2\pi\int\limits_0^{\infty} f_{\eta_\rho,0}^\pm(x_0,\rho)C(x_0,\rho)\rho d\rho  \ \ \ \ &m=0
\\
0                                            &m\neq 0,
\end{cases}
\end{equation}
where
$f_{\eta_\rho,0}^\pm(x_0,\rho)$   satisfies 
\begin{equation}																\label{eq:4.2}
Lf_{\eta_{\rho,0}}^\pm=\Big(\frac{\partial}{\partial x_0}+\frac{A(x_0)}{\rho}\frac{\partial}{\partial \rho}\Big)^2f_{\eta_\rho,0}^\pm  
-\frac{\partial^2 f_{\eta_{\rho,0}}^\pm(x_0,\rho)}{\partial  \rho^2}=0.
\end{equation}
Note  that  
$f_{\eta_{\rho,0}}^\pm$   has  the following  initial conditions   (cf.  (\ref{eq:2.2}),  (\ref{eq:2.3}))
\begin{align}																	\label{eq:4.3}
&f_{\eta_{\rho,0}}^\pm\big|_{x_0=0}=\gamma e^{i\rho \eta_\rho},
\\
\nonumber
&\frac{\partial f_{\eta_{\rho,0}}^\pm}{\partial x_0}\Big|_{x_0=0}
=i\lambda^\mp(\eta_\rho) \gamma e^{i\eta_\rho \rho},
\end{align}
where
\begin{equation}																\label{eq:4.4}
\lambda^\pm(\eta_\rho)=-\frac{A(0)}{\rho}\eta_\rho\pm\sqrt{\eta_\rho^2+1},\ \ \gamma  =\frac{1}{\sqrt 2\sqrt\rho}(\eta_\rho^2+1)^{-\frac{1}{4}}.
\end{equation}
Let 
\begin{equation}																\label{eq:4.5}
E=\gamma e^{-  i\eta_\rho\sigma(\rho,x_0)},\ \ \eta<0,
\end{equation}
where  
\begin{equation}																\label{eq:4.6}
\frac{\partial \sigma(\rho,x_0)}{\partial x_0}+\Big(\frac{A(x_0)}{\rho}+1\Big)\frac{\partial\sigma}{\partial\rho}=0,\ \ \sigma(\rho,0)=\rho.
\end{equation}
Note that $\sigma=\sigma(\rho,x_0),\sigma(\rho,0)=\rho$   is  the inverse  function   to
$\rho=\rho(\sigma,x_0)$  where  $\rho(\sigma,x_0)$  is the solution  of
\begin{equation}																\label{eq:4.7}
\frac{d\rho}{d x_0}=\frac{A(x_0)}{\rho}+1,\ \ \rho(\sigma,0)=\sigma.
\end{equation}
Indeed,   differentiating identity  $\sigma(\rho(\sigma,x_0),x_0)=\sigma$   in  $x_0$   we get
$$
\frac{\partial \sigma}{\partial  x_0}   +\frac{\partial \sigma}{\partial \rho}\Big(\frac{A}{\rho}+1\Big)=0.
$$
Note  that
\begin{equation}																\label{eq:4.8}
\frac{\partial}{\partial x_0}E\Big|_{x_0=0}= E( -i\eta_\rho)\frac{\partial}{\partial x_0}\sigma=
E\Big(i\eta_\rho
\frac{A}{\rho}\frac{\partial\sigma}{\partial\rho}+i\eta_\rho\frac{\partial\sigma}{\partial\rho}\Big)\Big|_{x_0=0}.
\end{equation}
Note that  $\frac{\partial\sigma}{\partial\rho}\Big|_{x_0=0}=1$.
Comparing  (\ref{eq:4.8})  and (\ref{eq:4.3})
we  have  that  factor  $\sqrt{\eta_\rho^2+1}$ in  (\ref{eq:4.3})    is  replaced  by  $|\eta_\rho|$  in  (\ref{eq:4.8}).

For the  simplicity  of notations we  shall  denote  $f_{\eta_\rho,0}^\pm$   for  $\eta_\rho<0$   by  $f_0^{+-}$  and we shall write  $\eta$  instead  of  $\eta_\rho$.

We shall construct  $f_0^{+-}$  in the form
\begin{equation}																	\label{eq:4.9}
f_0^{+-}=E +f_1^{+-} +f_2^{+-},
\end{equation}
where 
\begin{align}																		\label{eq:4.10}
&L(E + f_1^{+-})=0,
\\
\nonumber
&f_1^{+-}\big|_{x_0=0} = \frac{\partial f_1^{+-}}{\partial x_0}\Big|_{x_0=0}=0,
\end{align}
$L$  is the same as in  (\ref{eq:4.2})  and  $f_2^{+-}$  satisfies
\begin{align}																		\label{eq:4.11}
&Lf_2^\pm=0,
\\
\nonumber
&f_2^{+-} \big|_{x_0=0}=0,\  \frac{\partial f_2^{+-}}{\partial x_0}\Big|_{x_0=0}
=-i\gamma\big(\sqrt{\eta^2+1}-|\eta|\big)e^{-i\eta\rho}
=\gamma\frac{-i}{\sqrt{\eta^2+1}+|\eta|}e^{-i\eta\rho}.
\end{align}
We shall  estimate  $f_1^{+-}$   and  $f_2^{+-}$  later,  but  first  we  compute  the  contribution  to the  Hawking  radiation  of 
the  principal  term  $E$.

We shall define  the  wave  packet  $C_0(x_0,\rho)$  as
\begin{multline}																		\label{eq:4.12}
C_0(x_0,\rho)=\frac{1}{\sqrt\rho} e^{i\alpha\ln(\sigma(\rho,x_0)-\sigma^*)}
e^{-a(\sigma(\rho,x_0)-\sigma^*)}
\\ \cdot
(\sigma(\rho,x_0)-\sigma^*)^\e\theta(\sigma(\rho,x_0)-\sigma^*)
\end{multline}
where  $\sigma^*$   is the  same  as  in  (\ref{eq:3.5}).

There are following important features of  $C_0(x_0,\rho)$:    

a) when  $\rho\rw \rho_*$   then  $\sigma(\rho,x_0)\rw\sigma^*$  and vice versa  
since  $\sigma(\rho,x_0)$  is an inverse  function  to $\rho=\rho(\sigma,x_0)$.   Therefore,  $\mbox{supp}\, C_0(\rho,x_0)$  tends  to  
$\{\rho=\rho^*(x_0)\}$
when  $a\rw \infty$;

b) $\alpha\ln (\sigma(\rho,x_0)-\sigma*)$  is singular  on the event horizon  $\rho=\rho^*(x_0)$.   

To compute the average of the number of created  particles  we  need  to    
compute  (cf.   (\ref{eq:2.25}))
\begin{equation}																	\label{eq:4.13}
\int\limits_{-\infty}^\infty\big(|C^{-+}(\eta|^2-|C^{+-}(\eta)|^2\big)d\eta,
\end{equation}
where
\begin{equation}																	\label{eq:4.14}
C^{-+}(\eta)=-<f_0^{-+},C_0>\Big|_{x_0=0}=C_1^{-+}+C_2^{-+},
\end{equation}
\begin{equation} 																	\label{eq:4.15}
C^{+-}(\eta)=<f_0^{+-},C_0>\Big|_{x_0=0}=C_1^{+-}(\eta)+C_2^{+-}(\eta).
\end{equation}
We  have
\begin{equation}																	\label{eq:4.16}
C_1^{+-}(\eta)=i\int\limits_0^\infty\Big(\overline{f_0^{+-}}\frac{\partial C_0}{\partial x_0}
+\frac{A(x_0)}{\rho} \overline{f_0^{+-}}\frac{\partial C_0}{\partial \rho}\Big)\rho d\rho,
\end{equation}
\begin{equation}																	\label{eq:4.17}
C_2^{+-}=-i\int\limits_0^\infty\Big(\frac{\partial \overline{f_0^{+-}}}{\partial x_0} C_0
+\frac{A(x_0)}{\rho}\frac{\partial\overline{f_0^{+-}}}{\partial\rho}C_0\Big)\rho d\rho.
\end{equation}
Analogously,
we  define $C_1^{-+}(\eta)$  and  $C_2^{-+}(\eta)$.  We   first  consider  the  case  when  $f_0^{+-}$  is  replaced  by
$$
E=\frac{1}{\sqrt 2 \sqrt\rho(\eta^2+1)^{\frac{1}{4}}}\ e^{-i\eta\sigma(\rho,x_0)},  \eta<0.
$$
We have
\begin{equation}																	\label{eq:4.18}
C_{10}^{+-}(\eta)=\int\limits_0^\infty\frac{i}{\sqrt 2 (\eta^2+1)^{\frac{1}{4}}\sqrt \rho}e^{i\eta\sigma(\rho,x_0)}
\Big(\frac{\partial C_0}{\partial x_0}  
+\frac{A(x_0)}{\rho}\frac{\partial C_0}{\partial \rho}\Big)\rho d\rho,
\end{equation}
$\eta<0$,  where  $C_{10}^{+-}$   is the  same  as  $C_1^{+-}$   with  $f_0^{+-}$  replaced  by  $E$.
It follows from  (\ref{eq:4.12})  that 
$$																	
C_0=\frac{1}{\sqrt\rho}C_{01}(\sigma(\rho,x_0)) \ \mbox{where}\ \ C_{01}(\sigma)
=e^{i\alpha\ln (\sigma-\sigma_*)-a(\sigma-\sigma^*)}
(\sigma-\sigma_*)^\e \theta(\sigma-\sigma_*).
$$ 
Therefore,
\begin{equation}																	\label{eq:4.19}
\frac{\partial C_0}{\partial x_0}  
+\frac{A(x_0)}{\rho}\frac{\partial C_0}{\partial \rho}=
\frac{A(x_0)}{\rho}\Big(\frac{\partial}{\partial \rho}\frac{1}{\sqrt \rho}\Big)C_{01}(\sigma(\rho,x_0))-
\frac{1}{\sqrt\rho}\frac{\partial}{\partial \rho}C_{01}(\sigma(\rho,x_0)\Big)
\end{equation}
since
$$
\frac{\partial C_{01}}{\partial x_0}+\frac{A}{\rho}\frac{\partial C_{01}}{\partial \rho}+\frac{\partial C_{01}}{\partial\rho}=0.
$$
Thus
$$
C_{10}^{+-}=\hat C_{10}^{+-}+I_1,
$$
where
\begin{equation}																	\label{eq:4.20}
\hat C_{10}^{+-}=\int\limits_0^\infty\frac{i}{\sqrt 2 (\eta^2+1)^{\frac{1}{4}}}e^{i\eta\sigma(\rho,x_0)}\Big(-\frac{\partial}{\partial \rho}\Big)
C_{01}(\sigma(\rho,x_0))d\rho,
\end{equation}
\begin{equation}																	\label{eq:4.21}
I_1=\int\limits_{\rho^*(x)}^\infty\frac{A(x_0)}{\rho}\Big(\frac{\partial}{\partial \rho}\frac{1}{\sqrt\rho}\Big)
\frac{1}{\sqrt \rho}\frac{i}{\sqrt 2 (\eta^2+1)^{\frac{1}{4}}}
e^{i\eta\sigma(\rho,x_0)}C_{01}(\sigma(\rho,x_0)-\sigma_*)\rho d\rho.
\end{equation}
Making  in (\ref{eq:4.20})  change of variable  $\sigma=\sigma(\rho,x_0)$  we get
\begin{equation}																	\label{eq:4.22}
\hat C_{10}^{+-}=\int\limits_{\sigma_*}^\infty\frac{i}{\sqrt 2(\eta^2+1)^{\frac{1}{4}}} \ e^{i\eta\sigma}
\Big(-\frac{\partial}{\partial \sigma}\Big) C_{01}(\sigma-\sigma_*)d\sigma.
\end{equation}
Performing the Fourier  transform  in (\ref{eq:4.22})  and replacing  $\sigma-\sigma_*$  by  $\sigma'$   we get
\begin{equation}																	\label{eq:4.23}
\hat C_{10}^{+-}=e^{i\sigma_*\eta}\frac{i}{\sqrt 2(\eta^2+1)^{\frac{1}{4}}} i\eta\tilde C_{01}(-\eta),
\end{equation}
where
\begin{multline}																		\label{eq:4.24}
\tilde C_{01}(-\eta)=\int\limits_0^\infty e^{i\sigma'\eta}  e^{\alpha \ln\sigma'-a\sigma'}(\sigma')^\e d\sigma'=
\frac{\Gamma(i\alpha+\e+1)}{(\eta+ia)^{i\alpha+\e+1}}
e^{i\frac{\pi}{2}(i\alpha +\e+1)} 
\\
=e^{i\frac{\pi}{2}(i\alpha+\e)}\Gamma_0(i\alpha+\e+1)e^{(-i\alpha-\e -1)\ln(\eta+ia)} e^{i\frac{\pi}{2}(i\alpha+\e+1)}
\end{multline}
Here  (cf.  [2])
\begin{equation}																	\label{eq:4.25}
\Gamma(i\alpha+\e+1)
=e^{i\frac{\pi}{2}(i\alpha+\e)}\Gamma_0(i\alpha+\e+1),
\Gamma_0(i\alpha+\e+1) = i\int\limits_0^\infty  e^{(i\alpha+\e)\ln y-iy}dy.
\end{equation}
Since $\eta<0,a>0$,  we have  $\arg(\eta +ia)=\pi -\sin^{-1}\frac{a}{\sqrt{\eta^2+a^2}}$.

Analogously,
$$
C_{20}^{+-}=\hat C_{20}^{+-}-I_1,
$$
where
\begin{equation}																		\label{eq:4.26}
\hat C_{20}^{+-}=-i\int\limits_0^\infty\frac{e^{i\eta\sigma_*}}{\sqrt 2(\eta^2+1)^{\frac{1}{4}}}\Big(-\frac{\partial}{\partial\rho}e^{i\eta\sigma(\rho,x_0)}\Big)
C_{01}(\sigma(\rho,x_0))d\rho.
\end{equation}
Since  $\frac{\partial}{\partial \rho}  e^{i\eta\sigma(\rho,x_0)}=i\eta\frac{\partial\sigma}{\partial\rho}e^{i\eta\sigma(\rho,x_0)}$
we get,  changing variables  $\sigma=\sigma(\rho,x_0):$
\begin{equation}																		\label{eq:4.27}
\hat C_{20}^{+-}=\frac{ie^{i\eta\sigma_*}}{\sqrt 2(\eta^2+1)^{\frac{1}{4}} } i\eta \tilde C_{01}(-\eta).
\end{equation}
Analogously  to [2]  we have
\begin{equation}																		\label{eq:4.28}
C^{-+}=-\hat C_1^{+-}+\hat C_2^{+-},
\end{equation}
and,  therefore,   from  (\ref{eq:4.13}),   we get
\begin{equation}																		\label{eq:4.29}
|C^{-+}|^2-|C^{+-}|^2=-4\Re \hat C_1^{+-}\overline{\hat C_2^{+-}}.
\end{equation}
Thus,
\begin{equation}																		\label{eq:4.30}
\langle\Psi|N|\Psi\rangle=-\int\limits_0^\infty 4 \Re \hat C_1^{+-}\overline{\hat C_2^{+-}}d\eta.
\end{equation}
When
$f_0^{+-}$  is replaced  by  $E$    we  have,  as in [2], 
\begin{equation}																		\label{eq:4.31}
-\int\limits_0^\infty 4  \Re \hat C_{10}^{+-}\overline{\hat C_{20}^{+-}}d\eta=\int\limits_0^\infty
\frac{2\eta^2 |  \Gamma_0(i\alpha+\e+1)|^2 
e^{-2\alpha \sin^{-1}\frac{a}{\sqrt{\eta^2+a^2} }  } }
{\sqrt{\eta^2+1}|\eta+ia|^{2\e+2}} d\eta.
\end{equation}
Changing $\eta=\eta' a$  and  taking the limit  when  $a\rw\infty$
we  get
\begin{multline}																				\label{eq:4.32}
-\int\limits_0^\infty 4\Re  \hat C_{10}^{+-}\overline{\hat C_{20}^{+-}}d\eta
\\
=a^{-2\e}|\Gamma_0(i\alpha+\e+1)|^2\int\limits_0^\infty\frac{2\eta}{|\eta^2+1|^{\e+1}}
e^{-2\alpha \sin^{-1}\frac{1}{\sqrt{\eta^2+1}}}
 d\eta
+ O(a^{-2\e-1}).
\end{multline}
We used  in  (\ref{eq:4.32})  that  $\frac{\eta}
{(\eta^2+\frac{1}{a^2})^{\frac{1}{2}}   }
\rw 1$  when  $a\rw \infty$.

Now  we  shall  normalize  $C_0$.  We have (cf.  (3.8)  in [2])
\begin{multline}																				\label{eq:4.33}
<C_0,C_0>\int\limits_{\rho^*(x_0)}^\infty\int\limits_0^{2\pi}\Big(\frac{1}{\sqrt \rho}(\sigma(\rho,0)-\sigma_*)^\e  e^{-a\big(\sigma(\rho,0)-\sigma^*\big)}\Big)^2
\\
\cdot
\frac{2\alpha \frac{\partial\sigma}{\partial\rho}} {(\sigma(\rho,0)-\sigma_*)}\rho d\rho d\varphi
=4\pi\alpha \int\limits_{\sigma_*}^\infty(\sigma-\sigma_*)^{2\e-1}e^{-2a(\sigma-\sigma^*)}d\sigma.
\end{multline}
Note that  we  changed   variables  $\sigma=\sigma(\rho,\theta)$  in (\ref{eq:4.33}).
Therefore
\begin{equation}																			\label{eq:4.34}
<C_0,C_0>= 4\pi\alpha\int\limits_0^\infty \sigma^{2\e-1} e^{-2a\sigma}d\sigma =\frac{4\pi\alpha  \Gamma(2\e)}{(2a)^{2\e}}.
\end{equation}
 Denote
\begin{equation}																			\label{eq:4.35}
 C_n=\frac{C_0}{<C_0,C_0>^{\frac{1}{2}}    },\ \ 
 N_n(x_0)=\frac{N}{<C_0,C_0>}.
\end{equation}
Then
\begin{equation}																			\label{eq:4.36}
\langle\Psi | N_n|\Psi\rangle =\frac{\langle\Psi | N|\Psi\rangle}{<C_0,C_0>}.
\end{equation}
Thus   (cf.  [3])
\begin{equation}																			\label{eq:4.37}
\lim_{a\rw \infty}\langle\Psi | N_n|\Psi\rangle=\frac{2^{2\e}|\Gamma_0(i\alpha+\e+1)|^2}{2\pi\alpha\Gamma(2\e)}
\int\limits_0^\infty\frac{\eta}{|\eta^2+1|^{\e+1}}
e^{-2\alpha \sin^{-1}\frac{1}{\sqrt{\eta^2+1}}}
 d\eta,
\end{equation}
where we replaced  in  (\ref{eq:4.32})  and  (\ref{eq:4.37})  $f_0^{+-}$  by $E$.

Now  we shall  compute the contribution  of $f_1^{+-}$  and  $f_2^{+-}$,  and we will  show that they contribute  to the  lower order  
terms  in $\frac{1}{a}$,
thus  (\ref{eq:4.32})  holds  for  $f_0^{+-}$  and not only  for  $E^{+-}$.
In coordinates  $(\sigma,x_0)\ \ f_{1}^{+-} $  satisfies
\begin{equation}																				\label{eq:4.38}
Lf_{1}^{+-}\equiv \frac{\partial^2}{\partial x_0^2}f_{1}^{+-}+b_1\frac{\partial^2 f_{1}^{+-}}{\partial x_0\partial \sigma}
+b_2\frac{\partial f_{1}^{+-}}{\partial x_0} +b_3\frac{\partial f_{1}^{+-}}{\partial\sigma}
=-LE,
\end{equation}  
\begin{equation}																			\label{eq:4.39}
f_{1}^{+-}\Big|_{x_0=0}=\frac{\partial f_1^{+-}}{\partial x_0}\Big|_{x_0=0}=0.
\end{equation}
Note that 
$$
LE=b_3(x_0,\sigma) \gamma  (-i\eta)e^{-i\eta\sigma},
$$
where $\gamma$   is the same  as  in ({\ref{eq:4.4}).
Let  $G(x_0,\sigma,x_0',\sigma')$  be  the Green function for  $ L$
\begin{equation}																			\label{eq:4.40}
G\Big|_{x_0=0}=0,\ \ \ \frac{\partial G}{\partial x_0}\Big|_{x_0=0}=0.
\end{equation}
Then
\begin{equation}																			\label{eq:4.41}
f_{1}^{+-}=\int\limits\int\limits_{D(x_0,\sigma)} G(x_0,\sigma,x_0',\sigma')b_3(x_0',\sigma')\gamma (-i\eta)
\gamma 
e^{ -i\eta\sigma'}dx_0' d\sigma',
\end{equation}
where  $D(x_0,\sigma)$  is  the  demain  of  dependence  of  $(x_0,\sigma)$.

Integrating  by  parts  using  the identity  
\begin{equation}	\nonumber																		
e^{-i\sigma\eta}=-\frac{1}{i\eta}\frac{d}{d\sigma}e^{-i\eta\sigma}
\end{equation}
we get
\begin{equation}																			\label{eq:4.42}
|f_{1}^{+-}|\leq \frac{C}{1+|\eta|}\, \frac{1}{(1+\eta^2)^{\frac{1}{4}}},
\ \ \Big|\frac{\partial f_{1}^{+-}}{\partial\sigma}\Big|\leq \frac{C |\eta|}{1+|\eta|}\,\frac{1}{(1+\eta^2)^{\frac{1}{4}}}.
\end{equation}
Substituting  in (\ref{eq:4.13})  $f_{1}^{+-}$  instead  of  $f_0^{+-}$  we get  $C_{11}^{+-}$  and  $C_{12}^{+-}$,   where  $C_{11}^{+-}$
and $C_{12}^{+-}$  are  the same  as  in  (\ref{eq:4.16}),  (\ref{eq:4.17})  with   $f_0^{+-}$   replaced  by  $f_1^{+-}$.

Using  (\ref{eq:4.42})  we get
\begin{multline}																			\label{eq:4.43}
|C_{11}^{+-}|\leq\int\limits_0^\infty\frac{C}{(\eta^2+1)^{\frac{1}{4}}}\frac{1}{(1+|\eta|)}
\Big|\frac{\partial}{\partial\rho}C_{01}\Big|\frac{\partial\sigma}{\partial\rho}d\rho
\\
\leq C\int\limits_0^\infty\frac{1}{(\eta^2+1)^{\frac{1}{2}}(1+|\eta|)}\Big|\frac{\partial}{\partial\sigma}C_{01}(\sigma)\Big|d\sigma.
\end{multline}
Since  
$$
\big|\frac{\partial }{\partial \sigma} C_{01}(\sigma')\Big|\leq C e^{-a\sigma'}
\Big|\frac{\alpha}{\sigma'}+\frac{\e}{\sigma}+a\Big| \sigma^\e,
$$
we get,  changing variable $a\sigma=t$,
\begin{equation}																		\label{eq:4.44}
\big| C_{11}^{+-}\big|\leq  \frac{C}{|\eta_\rho^2+1|^{\frac{1}{4}}(|\eta|+1)}\frac{1}{a^{2\e}}.
\end{equation}
Analogously,  since  $|C_{01}(\sigma')|\leq  C e^{-a\sigma'}(\sigma')^\e$,  we  get
\begin{equation}																		\label{eq:4.45}
\big| C_{21}^{+-}\big|\leq  \frac{C|\eta|}{|\eta_\rho^2+1|^{\frac{1}{4}}(|\eta|+1)}\frac{1}{a^{\e+1}}.
\end{equation}
Therefore
\begin{equation}																		\label{eq:4.46}
-\int\limits_0^\infty 4\Re C_{11}^{+-}\,\overline{C_{21}^{+-}}d\eta   \leq C\int\limits_0^\infty \frac{1}{(|\eta|+1)^2}\,\frac{1}{a^\e}\, \frac{1}{a^{\e+1}}
\leq \frac{C}{a^{2\e+1}}.
\end{equation} 
Thus  $f_1^{+-}$  contributes    to a higher  order  term  in  $\frac{1}{a}$.

Now  we shall  estimate  the contribution  of  $f_2^{+-}$  where  $f_2^{+-}$  is the same  as  in  (\ref{eq:4.11}),  i.e.  
$$
f_2^{+-}\big|_{x_0=0}=0,\ \frac{\partial f_2^{+-}}{\partial x_0}\big|_{x_0=0}=\frac{-i\gamma e^{-i\eta\rho}}{\sqrt{\eta^2+1}+|\eta|}.
$$
Note  that  $\sigma(\rho,0)=\rho$.  We have,  in  $(\sigma,x_0)$  coordinates:
\begin{equation}	     																	\label{eq:4.47}
f_2^{+-} =\frac{-i\gamma}{\sqrt{\eta^2+1}+|\eta|} x_0\, e^{-i\eta\sigma}  +f_3^{+-},
\end{equation}
where
\begin{equation}																		\label{eq:4.48}
f_3^{+-}\big|_{x_0=0}=0,\ \frac{\partial f_3^{+-}}{\partial x_0}\big|_{x_0=0}=0,
\end{equation}
\begin{multline}																		        \label{eq:4.49}
Lf_3^{+-} =-L\Big(\frac{-i\gamma  x_0 e^{-i\eta\sigma}}{\sqrt{\eta^2+1}+|\eta|}\Big)
\\
=\frac{\gamma \eta}{\sqrt{\eta^2+1}+|\eta|}b_1e^{-i\eta\sigma}
-\frac{b_3\gamma x_0 \eta e^{-i\sigma\eta}}{\sqrt{\eta^2+1}+\eta}
+\frac{ib_2\gamma  e^{-i\eta\sigma}}{\sqrt{\eta^2+1}+|\eta|}.
\end{multline}
Thus,   $f_3^{+-}$  satisfies  the same    estimates  as $f_2^{+-}$.  Therefore,
\begin{equation}																		\label{eq:4.50}  
|f_3^{+-}|\leq  \frac{C}{(\eta^2+1)^{\frac{1}{4}}(\sqrt{\eta^2+1}+|\eta|)},\ \ 
\Big|\frac{\partial f_3^\pm}{\partial \sigma}\Big| \leq 
\frac{C|\eta|}{(\eta^2+1)^{\frac{1}{4}}(\sqrt{\eta^2+1}+|\eta|)}.
\end{equation}
Having estimates (\ref{eq:4.50})  we can prove,  as for $f_1^{+-}$,  
that
\begin{equation}																		\label{eq:4.51}
|C_{13}|\leq  \frac{C}{(\eta^2+1)^{\frac{1}{4}}(\sqrt{\eta^2+1}+|\eta|)}
\frac{1}{a^{\e+1}},
\end{equation}
\begin{equation}																		\label{eq:4.52}
|C_{23}|\leq \frac{C |\eta|}{(\eta^2+1)^{\frac{1}{4}}(\sqrt{\eta^2+1}+|\eta|)}
\frac{1}{a^{\e+1}}.
\end{equation}
Therefore  we proved
\begin{theorem}																		\label{theo:4.1}
 Let $C_0$  be the same as in  (\ref{eq:4.12})  and $ C_n$  be the same as in (\ref{eq:4.33}),  (\ref{eq:4.35}).
 Then
\begin{equation}																		\label{eq:4.53}
\lim_{a\rw \infty} \langle \Psi | N_n|\Psi\rangle =
\frac{2^\e|\Gamma_0(i\alpha+\e+1)|^2}{2\pi \alpha\Gamma(2\e)}
\int\limits_0^\infty
\frac
{\eta e^{   - 2\sin^{-1}\frac{1}{\sqrt{\eta^2+1}} }
}
{(\eta^2+1)^{\e+1}}d\eta.
\end{equation}
\end{theorem}

\end{document}